\documentclass[11pt,a4paper,twoside,groupcitations]{article}
\usepackage[T1]{fontenc}
\usepackage[ansinew]{inputenc}
\usepackage[english]{babel}
\usepackage{amsfonts}
\usepackage{amsmath}
\usepackage{bm}
\usepackage{array}
\usepackage{amsthm}
\usepackage{amssymb}
\usepackage{graphicx}
\usepackage{subfigure}
\usepackage{braket}
\usepackage{eucal}
\usepackage{verbatim}
\usepackage[table]{xcolor}
\usepackage{caption}
\usepackage{cite}
\usepackage{textcomp}
\usepackage{multicol}
\usepackage{tikz}
\usetikzlibrary{positioning,arrows}
\usetikzlibrary{decorations.pathmorphing}
\usetikzlibrary{decorations.markings}
\usetikzlibrary{calc,decorations.markings}
\usetikzlibrary{arrows,shapes}
\usetikzlibrary{matrix,arrows}
\usepackage{pgfplots}
\usepackage{xparse}
\definecolor{jade}{HTML}{00A86B}
\newcommand{\be}{\begin{eqnarray}}
\newcommand{\ee}{\end{eqnarray}}

\newcommand{\expec}[1]{\mbox{$\langle\, #1\,\rangle$}}

\renewcommand{\d}{\mbox{${\rm d}$}} 
\newcommand{\lp}{\ell_{\rm p}}

\newcommand{\gn}{G_{\rm N}}
\newcommand{\rh}{r_{\rm H}}

\title{\bf On gravitational collapse and integrable singularities}
\author{Roberto~Casadio$^{abc}$\thanks{E-mail: casadio@bo.infn.it}
$\,$,
Andrea~Giusti$^{abc}$\thanks{E-mail: andrea.giusti9@unibo.it}
$\,$,
Alexander~Kamenshchik$^{ab}$\thanks{E-mail: kamenshchik@bo.infn.it}
$\,$
and
Jorge~Ovalle$^{de}$\thanks{E-mail: jorge.ovalle@physics.slu.cz}
\\
\\
$^a${\em Dipartimento di Fisica e Astronomia, Universit\`a di Bologna}
\\
{\em via Irnerio~46, 40126 Bologna, Italy}
\\
\\
$^b${\em I.N.F.N., Sezione di Bologna, I.S.~FLAG}
\\
{\em viale B.~Pichat~6/2, 40127 Bologna, Italy}
\\
\\
$^c${\em Alma Mater Research Center on Applied Mathematics (AM$^2$)}
\\
{\em Via Saragozza 8, 40123 Bologna, Italy}
\\
\\
$^d${\em Research Centre for Theoretical Physics and Astrophysics, Institute of Physics}
\\
{\em Silesian University in Opava, CZ-746 01 Opava, Czech Republic}
\\
\\
$^e${\em Universidad de Tarapac\'a}
\\
{\em Avenida Luis Emilio Recabarren 2477, Iquique, Chile}
}
\begin{document}
\maketitle
\begin{abstract}
Schwarzschild black holes are expected to emerge as the end states of the
classical gravitational collapse from non-singular configurations.
After integrable curvature singularities appear, the interior geometry can be modelled
to exhibit a transition, called ``Minkowski breaking'', when the inner horizon disappears,
before all matter collapses into the central singularity.
This picture implies a quantum framework to describe the final stages of the gravitational collapse,
and here we will provide more insights from the semiclassical approximation for the
energy-momentum tensor and the Madelung approximation for collapsing matter.
In particular, we will show that the quantum potential in the Raychaudhuri equation
starts to strongly oppose the collapse towards the Schwarzschild singularity precisely after
the Minkowski breaking.
\end{abstract}
\section{Introduction}
\setcounter{equation}{0}
\label{S:intro}
One of the main challenges of contemporary astrophysics is to understand the gravitational collapse
of realistic matter towards the formation of objects that should be described by classical black hole
geometries in the Einstein theory of gravity~\cite{HE}.
A crucial feature of such geometries is the presence of singularities that would correspond to diverging
tidal forces acting on collapsing matter.
Such mathematical divergences in physical models typically indicate the need for a better description
of the process.
\par
The singularities of known black hole solutions of the Einstein equations~\cite{HE} can be removed
by imposing regularity conditions on scalar invariants and assuming that the (effective) energy density
$\rho\sim r^0$ for $r\to 0$.~\footnote{We limit our discussion to spherically symmetric systems for
which the radial coordinate $r$ is usually assumed to be the areal radius.}
Such conditions are inspired by classical physics~\cite{Carballo-Rubio:2023mvr} and
result in an (effective) Misner-Sharp-Hernandez (MSH) mass
function~\cite{Misner:1964je,Hernandez:1966zia} satisfying
\be
m(r)
\equiv
4\,\pi
\int_0^r
\rho(x)\,x^2\,\d x
\sim
r^3
\ ,
\quad
{\rm for}
\ r\to 0
\ .
\label{Ccond}
\ee
However, the above MSH mass usually induces the appearance of an inner Cauchy horizon
(if the outer horizon has formed), which in turn breaks the global hyperbolicity of the spacetime
and triggers a potentially fatal instability known as mass inflation~\cite{Poisson:1989zz}.
\par
A different framework can be implemented based on the possibility
that black hole interiors and the collapsed matter therein are described
more accurately by quantum physics
(see, {\em e.g.}~Refs.~\cite{Saini:2014qpa,Greenwood:2008ht,Wang:2009ay,Brustein:2015sma,
Davidson:2014tda,Dvali:2011aa,DeLorenzo:2014pta,Corda:2012dw,Calmet:2021cip}).
A simple argument along those lines was considered in Ref.~\cite{Casadio:2023iqt}:
in the hydrodynamic Madelung approximation of quantum mechanics~\cite{Madelung:1927ksh},
one can assume that the effective energy density $\rho\simeq \mu\,|\psi|^2$, where $\psi=\psi(r)$
is the collective wavefunction of the matter source and $\mu$ a mass.
This yields 
\be
m(r)
\simeq
4\,\pi
\int_0^r
\mu\,
|\psi(x)|^2\,x^2\,\d x
<
\infty
\qquad
{\rm for}
\ r>0
\ ,
\label{Qcond}
\ee
which accommodates for $\rho\sim r^{n-2}$ and $m\sim r^{1+n}$ with
$-1 < n <2$.
This behaviour near the centre of the collapsing object ensures that $m(0)=0$ and replaces
the central singularity with an integrable singularity,
that is a region where the curvature invariants and the effective energy-momentum tensor diverge
but their volume integrals remain finite~\cite{Lukash:2013ts}.
Additionally, no inner horizon is present for $-1<n\le 0$~\cite{Casadio:2021eio,Casadio:2022ndh}
and the corresponding instabilities should be avoided.
\par
The models of gravitational collapse proposed in Refs.~\cite{Ovalle:2025pue,Ovalle:2026lxb}
precisely display the emergence of the above (potentially quantum) behaviour~\eqref{Qcond},
starting from regular configurations with the MSH mass function satisfying the classical condition~\eqref{Ccond}.
This transition is described by a power $n$ that decreases along the collapse from values
corresponding to regular distributions ($n>2$) to integrable singularities ($-1<n<2$).
The interior containing an integrable singularity should then continue to collapse and the Cauchy horizon
suddenly disappear for $n=0$, a discontinuity which was termed ``Minkowski breaking''
in Ref.~\cite{Ovalle:2026lxb}. 
In this work, we review the simplest of those models and analyse in more details the effective
energy-momentum tensor in the semiclassical approximation.
One of the main results is that the Schwarzschild singularity can only be approached asymptotically
(after the Minkowski breaking).
Moreover, we will investigate the possibility that the collapse ends in a final (static) state 
before the Schwarzschild singularity forms by computing the contribution from the quantum potential
of the Madelung approximation that appears in the Raychaudhuri equation~\cite{Das:2013oda}
(see also Ref.~\cite{Pinto-Neto:2013toa} for the cosmological case).
Remarkably, the effect of this contribution changes drastically at the Minkowski breaking, after
which it starts to strongly oppose the collapse.
\par
In Section~\ref{S:model} we briefly review the main feature of the model~\cite{Ovalle:2025pue,Ovalle:2026lxb}
for which we then compute and analyse the effective energy-momentum tensor in Section~\ref{S:emt};
the Raychaudhuri equation and Madelung approximation for matter in a possibly static 
final state are then studied in Section~\ref{S:CQcoll}, with final remarks and outlook provided in Section~\ref{S:conc}.
\section{Modelling the collapse}
\setcounter{equation}{0}
\label{S:model}
Following previous works~\cite{Ovalle:2025pue,Ovalle:2026lxb,Ovalle:2024wtv}, we analyse the
dynamical evolution of the gravitational collapse from the moment the outer region is precisely
described by the vacuum Schwarzschild geometry~\cite{Schwarzschild:1916uq} with horizon
radius $\rh=2\,\gn\,M$, where $M$ is the Arnowitt-Deser-Misner (ADM)
mass~\cite{Arnowitt:1959ah} of the system.
This model is described by the metric 
\begin{equation}
\d s^{2}
=
g_{\mu\nu}\,\d x^\mu\,\d x^\nu
=
-f(v,r)\,\d v^{2}
+
2\,\d v\,\d r
+r^2\,d\Omega^2
\ ,
\label{EF-metric}
\end{equation}
where $v$ is the ingoing null coordinate related to the Schwarzschild time $t$ by $\d v=\d t+\d r/f$,
$d\Omega^2=\d\vartheta^2+\sin^2\vartheta\,\d\phi^2$, and 
\begin{equation}
f
=
\left\{
\begin{array}{ll}
1-\strut\displaystyle{\frac{2\,\gn\,m(v,r)}{r}}
\ ,
&
{\rm for}\
0< r \leq \rh
\\
\\
1-\strut\displaystyle{\frac{\rh}{r}}
\ ,
&
{\rm for}\
r>\rh
\ .
\end{array}
\right.
\label{eq:f}
\end{equation}
The metric function $f$ and its derivative $f'=\partial_r f$ are assumed to be continuous across $r=\rh$,
with $m(v,\rh)=M$ and $m'(v,\rh)=0$,
which yields the interior MSH mass~\cite{Ovalle:2025pue,Ovalle:2026lxb,Ovalle:2024wtv}
\begin{equation}
m
=
\frac{r}{2\,\gn\,[n(v)-2]}
\left\{
\frac{r^2}{\rh^2}
\left[n(v)+1\right]
-
3\left(\frac{r}{\rh}\right)^{n(v)}
\right\}
\ ,
\label{m-n}
\end{equation}
where $n=n(v)$ therefore entails the evolution~\cite{Ovalle:2026lxb}.
\par
\begin{figure}[t]
\centering
\includegraphics[height=0.30\textwidth]{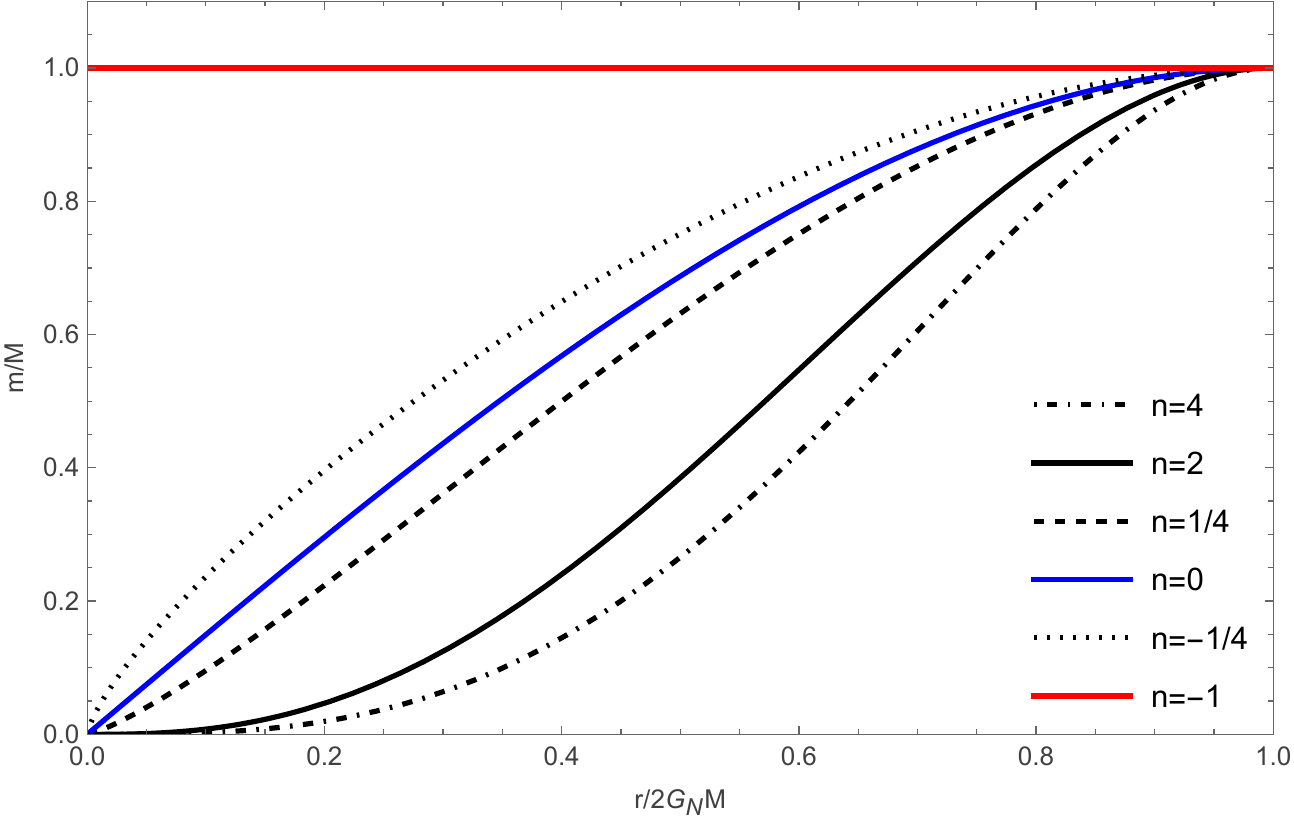}
$\ $
\includegraphics[height=0.30\textwidth]{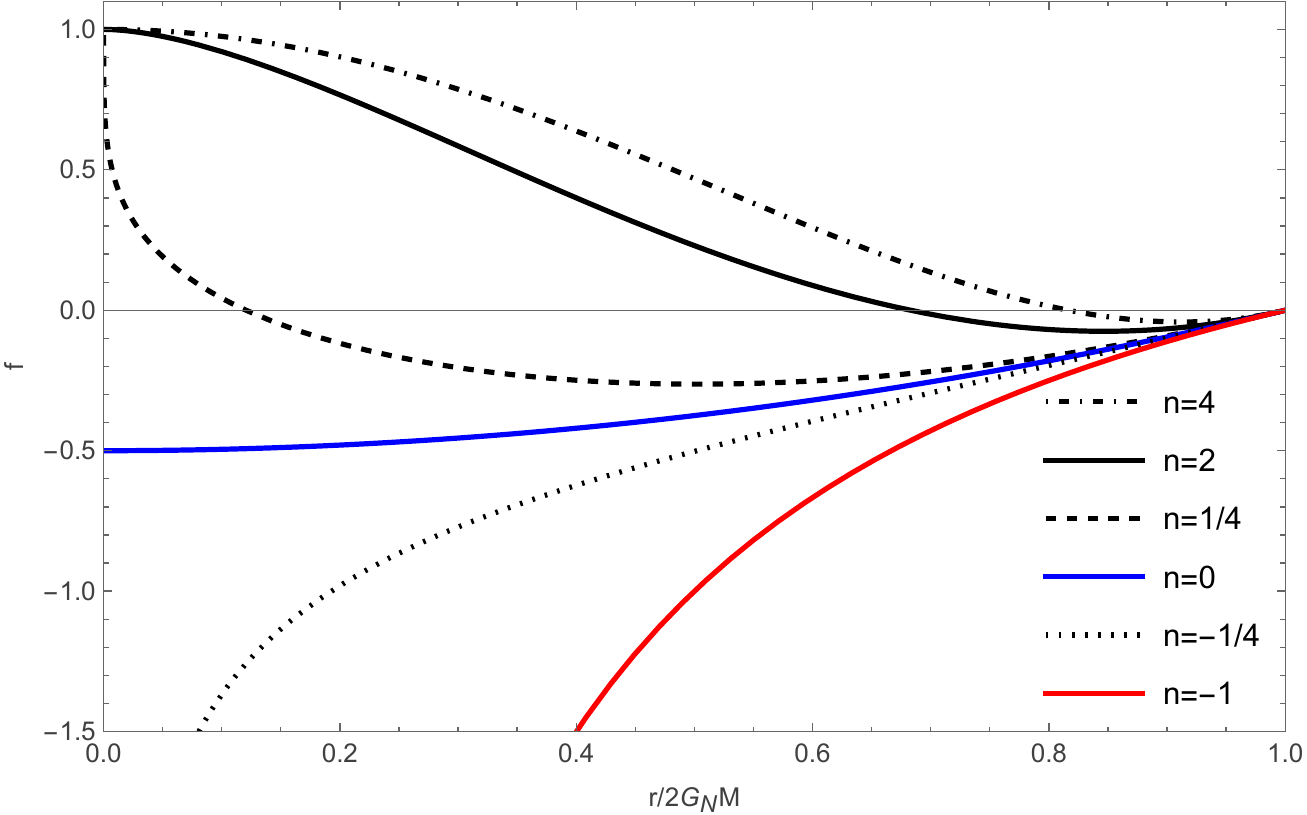} 
\caption{Left panel:
evolution of the mass function~\eqref{m-n} from a regular black hole with $n>2$ towards
the point-like (Schwarzschild) singularity for $n\to -1$.
Notice how the slope at $r=0$ jumps from zero to positive infinity across $n=0$ [see Eq.~\eqref{dm0}].  
Right panel: evolution of the metric function~\eqref{eq:f} for the cases in the left panel.
The inner horizon disappears for $n=0$ [see Eqs.~\eqref{jump} and \eqref{f0r}] across which $f(0)=1$
jumps to $f(0)\to -\infty$.
The event horizon is fixed at $r=\rh=2\,\gn\,M$.}
\label{figMF}
\end{figure}

The null convergence condition $R_{\mu\nu}\,l^\mu\,l^\nu\ge 0$, for any null vector $l^\mu\,l_\mu=0$,
enforces $\dot{n}\equiv \partial_v n < 0$ and the formation of singularities is unavoidable within the classical
dynamics~\cite{Ovalle:2025pue}.
This behaviour is clearly illustrated in Fig.~\ref{figMF}, which shows snapshots of the possible 
evolution beginning from an initial configuration with $n>2$, which is regular in the sense of
Eq.~\eqref{Ccond} (curvature scalars remain finite at $r=0$), that is
\be
m
\sim
r
\left(\frac{r}{\rh}\right)^2
\ ,
\quad
{\rm for}\
r\to 0
\ .
\ee
For $n=2$, we have
\be
m
\sim
\left(\frac{r}{\rh}\right)^2
r\,
\ln\left(\frac{\rh}{r}\right)
\ ,
\ee
and, for $-1\le n< 2$, we recover the condition~\eqref{Qcond} with
\be
m
\sim
r
\left(\frac{r}{\rh}\right)^n
\ ,
\quad
{\rm for}\
r\to 0
\ ,
\ee
so that the interior contains an integrable singularity~\cite{Lukash:2013ts,Casadio:2023iqt},
approaching the Schwarzschild limit for $n\to -1$~\cite{Ovalle:2026lxb}.
\par
During this last stage of the process, a discontinuity arises that is characterised by the behaviour
of the derivative of the mass function in the centre,
\be
m'(v,0)
=
\left\{
\begin{array}{ll}
0
\ ,
&
{\rm for}\
n>0
\\
\\
1
\ ,
&
{\rm for}\
n=0
\\
\\
+\infty
\ ,
&
{\rm for}\
-1<n<0
\ ,
\end{array}
\right.
\label{dm0}
\ee
corresponding to
\begin{equation}
\label{jump}
f(v,0)=\left\{
\begin{array}{ll}
1
\ ,
&
{\rm for}
\
n>0
\\
\\
-1/2
\ ,
&
{\rm for}
\
n
=
0
\end{array}
\right.
\end{equation}
and 
\be
f(v,r\to 0)
\sim
-r^{-|n|}
\ ,
\quad
{\rm for}\
-1\le n<0
\ ,
\label{f0r}
\ee
with $n=-1$ reducing precisely to the Schwarzschild geometry with $m=M$ everywhere. 
Note that, since $f(\rh)=0$, there must exist a inner Cauchy horizon at $0<r_{\rm c}<\rh$,
where $f(r_{\rm c})=0$, for $n>0$.
\par
The discontinuities at $n=2$ and $n=0$ can be further analysed by noting that 
\be
f'(v,0)
=
\left\{
\begin{array}{ll}
0
\ ,
&
{\rm for}\
n
\ge
2
\\
\\
-\infty
\ ,
&
{\rm for}\
0<n<2
\\
\\
0
\ ,
&
{\rm for}\
n
=
0
\\
\\
+\infty
\ ,
&
{\rm for}\
-1<n<0
\ .
\end{array}
\right.
\label{dm0}
\ee
Moreover, the local minimum of the metric function $f$ is located at
\begin{equation}
\label{extrema2}
r_{\rm e}(v)
=
\rh
\left[\frac{2}{3}\left(1+\frac{1}{n(v)}\right)\right]^{\frac{1}{n(v)-2}}
\ .
\end{equation}
As $n(v)\to 0$, this extremum shifts toward the origin $r=0$ and causes the Cauchy horizon $r_{\rm c}$
to shrink correspondingly, since $r_{\rm c}(v)\le r_{\rm e}(v)\to 0$ for $n(v) \to 0$.
One therefore sees that, as $n(v)$ crosses $n=0$, the Cauchy horizon disappears.
This discontinuous transition was termed ``Minkowski breaking'' in Ref.~\cite{Ovalle:2026lxb}.
\par
The total ADM mass $M$, which is initially distributed throughout the whole interior region, 
also exhibits a discontinuous behaviour,
\begin{equation}
\label{jump2}
m(v,0)=\left\{
\begin{array}{ll}
0
\ ,
&
{\rm for}\ 
n(v)>-1
\ ,
\\
\\
M
\ ,
&
{\rm for}\
n(v)=-1
\ .
\end{array}
\right.
\end{equation}
This shows that the collapse of the total mass $M$ into the point-like Schwarzschild 
singularity at $r=0$ cannot occur continuously.~\footnote{It is worth recalling that General
Relativity does not admit Dirac delta-like sources~\cite{Geroch:1986jjl,tangherlini}.
\label{foot1}}
This will be further clarified by the analysis of the effective energy-momentum tensor
in the next section.
\section{Effective energy-momentum tensor}
\setcounter{equation}{0}
\label{S:emt}
The interior metric~\eqref{EF-metric} with the mass function~\eqref{m-n} for $0<r<\rh$, is
a generalised Vaidya metric~\cite{Wang:1998qx}, whose effective energy-momentum
tensor can be obtained from the Einstein tensor $G_{\mu\nu}$ and reads~\footnote{We
recall that mixed components $T^\mu_{\ \nu}$ and $G^\mu_{\ \nu}$ need not be symmetric.}  
\be
T^\mu_{\ \nu}
=
\frac{G^\mu_{\ \nu}}{8\,\pi\,\gn}
=
\left[
\begin{array}{cccc}
-\frac{m'}{4\,\pi\,r^2}
& 0 & 0 & 0
\\
\frac{\dot m}{4\,\pi\,r}
&
-\frac{m'}{4\,\pi\,r^2}
& 0 & 0
\\
0 & 0 &
-\frac{m''}{8\,\pi\,r}
&
0
\\
0 & 0 & 0 &
-\frac{m''}{8\,\pi\,r}
\end{array}
\right]
\ .
\label{Ta_b}
\ee
This procedure ensures that the energy-momentum tensor~\eqref{Ta_b} is covariantly
conserved,~\footnote{Furthermore, it is of type~II in the classification of Ref.~\cite{HE}.}
namely
\be
\nabla_\mu T^\mu_{\ \nu}
=
0
\ .
\ee
\par
The physical content of the energy-momentum tensor~\eqref{Ta_b} can be clarified
using suitable tetrads.
We will start from the usual null vectors~\cite{Wang:1998qx}
\be
l_\mu
=
\left(1,0,0,0\right)
\ ,
\qquad
n_\mu
=
\left(f/2,-1,0,0\right)
\ ,
\ee
satisfying
\be
l_\mu\,l^\mu
=
n_\mu\,n^\mu
=0
\ ,
\qquad
l_\mu\,n^\mu
=
-1
\ ,
\ee
using which we find that 
\be
T_{\mu\nu}
=
\left(\rho+p_{\rm t}\right)
\left(l_\mu\,n_\nu
+
l_\nu\,n_\mu\right)
+
\Phi\,l_\mu\,l_\nu
+
p_{\rm t}\,g_{\mu\nu}
\ ,
\label{Tij}
\ee
where the effective energy density and pressures are respectively given by
\be
\rho
=
\frac{m'}{4\,\pi\,r^2}
=
-p_{\rm r}
\ ,
\qquad
p_{\rm t}
=
-\frac{m''}{8\,\pi\,r}
\ ,
\label{rhoM}
\ee
and the flux term
\be
\Phi
=
\frac{\dot m}{4\,\pi\,r^2}
\ .
\label{eq:Phi}
\ee
\par
Orthonormal tetrads satisfying
\be
e^\mu_{(a)}\,g_{\mu\nu}\,e^\nu_{(b)}
=
\eta_{(a)(b)}
\equiv
{\rm diag}\left[-1,1,1,1\right]
\ ,
\ee
are then given by
\be
e^\mu_{(0)}
=
\frac{l^\mu+n^\mu}{\sqrt{2}}
\ ,
\qquad
e^\mu_{(1)}
=
\frac{l^\mu-n^\mu}{\sqrt{2}}
\ ,
\label{e^mu}
\ee
and the usual
\be
e^\mu_{(2)}
=
\left(
0,0,1/r,0
\right)
\ ,
\qquad
e^\mu_{(3)}
=
\left(
0,0,1/(r\,\sin\vartheta),0
\right)
\ .
\ee
\par
By projecting Eq.~\eqref{Tij} on this tetrad, we then find
\be
T_{(a)(b)}
\equiv
e^\mu_{(a)}\,T_{\mu\nu}\,e^\nu_{(b)}
=
\left[
\begin{array}{cccc}
\rho+\Phi/2 & \Phi/2 & 0 & 0
\\
\Phi/2 & p_{\rm r}+\Phi/2 & 0 & 0
\\
0 & 0 & p_{\rm t} & 0
\\
0 & 0 & 0 & p_{\rm t}
\end{array}
\right]
\ .
\ee
\par
Using the mass function in Eq.~\eqref{m-n}, we can next show the explicit expressions
for the effective energy density and radial pressure
\be
\rho
=
-p_{\rm r}
=
\frac{3}{8\,\pi\,\gn\,\rh^2}
\left(\frac{n+1}{n-2}\right)
\left[
1-
\left(\frac{r}{\rh}\right)^{n-2}
\right]
\ ,
\label{rhop_r}
\ee
the tension
\be
p_{\rm t}
=
\frac{3}{8\,\pi\,\gn\,\rh^2}
\left(\frac{n+1}{2-n}\right)
\left[
1-
\frac{n}{2}
\left(\frac{r}{\rh}\right)^{n-2}
\right]
\ ,
\label{p_t}
\ee
and the flux term
\be
\Phi
=
-\frac{3\,r\,\dot n}{8\,\pi\,\gn\,\rh^2\,(n-2)^2}
\left\{
1
-
\left(\frac{r}{\rh}\right)^{n-2}
\left[
1
+
(n-2)\,\ln\left(\frac{\rh}{r}\right)
\right]
\right\}
\ .
\label{Phi}
\ee
Note that on the horizon we have $\rho(\rh)=p_{\rm r}(\rh)=\Phi(\rh)=0$, with a tension
\be
p_{\rm t}(\rh)
=
\frac{3\,(n+1)}{16\,\pi\,\gn\,\rh^2}
\ ,
\ee
which supports the static horizon (for $n>-1$) and vanishes in the Schwarzschild limit $n\to -1^+$.
Note that the above solutions satisfy, at least, the null energy condition (as long as there is collapse),
that is when $\dot n<0$ (for more details, see Ref.~\cite{Ovalle:2025pue}).
\par
We can next study the above quantities near $r=0$ in the two regimes identified previously.
\subsection{Regular interior $n\ge 2$}
\label{SS:reg}
The leading order term of the density and pressures~\eqref{rhop_r} and \eqref{p_t}
near $r=0$ for $n>2$ is given by
\be
\rho
\sim
\frac{3}{8\,\pi\,\gn\,\rh^2}
\left(\frac{n+1}{n-2}\right)
\sim
-p_{\rm r}
\sim
-p_{\rm t}
>
0
\ ,
\ee
which shows a quasi-de~Sitter behaviour.
The flux~\eqref{Phi} also behaves as
\be
\Phi
\sim
-\frac{3\,r\,\dot n}{8\,\pi\,\gn\,\rh^2\,(n-2)^2}
\ .
\ee
\par
In the limit $n\to 2$, we also find
\be
\rho
\sim
\frac{9}{8\,\pi\,\gn\,\rh^2}\,
\ln\left(\frac{\rh}{r}\right)
\sim
-p_{\rm r}
\sim
-p_{\rm t}
>
0
\ ,
\ee
and
\be
\Phi
\sim
-\frac{3\,r\,\dot n}{16\,\pi\,\gn\,\rh^2}
\left[\ln\left(\frac{\rh}{r}\right)\right]^2
\ .
\ee
In all cases, density, radial pressure and tension integrated on balls centred around $r=0$ 
yield finite values that vanish for vanishing radius, like does the flux integrated on the surface of such balls,
\be
4\,\pi\,r^2\,\rho\,\d r
\sim
4\,\pi\,r^2\,\Phi
\sim
\left\{
\begin{array}{ll}
r^3
\ ,
&
{\rm for}
\ 
n>2
\\
\\
r^3\,\ln(r)
\ ,
&
{\rm for}
\ 
n=2
\ ,
\end{array}
\right.
\ee
as one expects from the covariant conservation of the energy-momentum tensor.
\subsection{Integrable interior $-1< n<2$}
\label{SS:sing}
The leading order term of the density and radial pressure~\eqref{rhop_r} near $r=0$ for $-1< n<2$
is given by
\be
\rho
\sim
\frac{3}{8\,\pi\,\gn\,\rh^2}
\left(\frac{n+1}{2-n}\right)
\left(\frac{\rh}{r}\right)^{2-n}
\sim
-p_{\rm r}
>
0
\ ,
\ee
the tension~\eqref{p_t} by
\be
p_{\rm t}
=
-\frac{3\,n}{15\,\pi\,\gn\,\rh^2}
\left(\frac{n+1}{2-n}\right)
\left(\frac{\rh}{r}\right)^{2-n}
\ ,
\ee
so that $p_{\rm t}<0$ for $0\le n<2$ and $p_{\rm t}>0$ for $-1<n<0$,
and the flux term~\eqref{Phi} by
\be
\Phi
=
-\frac{3\,r\,\dot n}{8\,\pi\,\gn\,\rh^2\,(2-n)}
\left(\frac{\rh}{r}\right)^{2-n}
\ln\left(\frac{\rh}{r}\right)
\ .
\ee
All of the above quantities now diverge for $r\to 0$ but the integral of density, pressure and
tension on balls centred around $r=0$ remains finite and vanishes for vanishing radius,
\be
4\,\pi\,r^2\,\rho\,\d r
\sim
4\,\pi\,r^2\,p_{\rm t}\,\d r
\sim
r^{1+n}
\ ,
\ee
like the flux integrated on the surface of such balls also remains well behaved,
\be
4\,\pi\,r^2\,\Phi
\sim
\dot n\,
r^{1+n}\,\ln(r)
\ .
\ee
\par
In particular, nothing appears particularly singular for $n=0$.
In fact, the only leading term that changes is the one for the tension~\eqref{p_t}, namely
\be
p_{\rm t}
\sim
-\frac{3}{16\,\pi\,\gn\,\rh^2}
<
0
\ ,
\ee
which becomes regular.
The Minkowski breaking occurring for $n\to 0^+$ appears instead associated with
the derivative of the geometric function
\begin{equation}
\dot f
\sim
\dot n\,r^{n}\,\ln(r)
\ ,
\end{equation}
for which a physical meaning is less obvious.
\par
For $n\to -1^+$, we further notice that 
\be
4\,\pi\,r^2\,\rho\,\d r
\sim
(n+1)\,
\d r/r
\ee
and
\be
4\,\pi\,r^2\,\Phi
\sim
\dot n\,\ln(r)
\ .
\ee
This yields a physical condition for the formation of the Schwarzschild singularity:
the proper energy flux (measured by a locally inertial observer) diverges in the centre
unless $\dot n(v_{\rm c})=0$ when $n(v_{\rm c})\to -1^+$.
One might therefore conclude that the Schwarzschild singularity can only be asymptotically
approached (in agreement with the comment in footnote~\ref{foot1}).
\section{Classical and quantum end state}
\setcounter{equation}{0}
\label{S:CQcoll}
So far we have not assumed any specific dynamics [encoded by the unspecified $n=n(v)$]
and found no indication that the collapse should stop before the Schwarzschild singularity
is approached for $n(v)\to -1^+$.
We will now study this question more in details, namely we will investigate whether there could
be (final) static configurations with $-1<n(v)\equiv n_{\rm s}<2$ and $\dot n(v)=0$ for $v\ge v_{\rm s}$.
\subsection{Geodesic motion}
The vector of components $e_{(0)}^\mu$ in Eq.~\eqref{e^mu} does not satisfy the geodesic equation,
since
\be
e_{(0)}^\mu\nabla_\mu
e_{(0)}^\nu
=
\frac{f'}{8}
\left(
2,2+f,0,0
\right)
\ .
\ee
We can then consider a (local) Lorentz transformation of the tetrad vectors~\eqref{e^mu},~\footnote{This
procedure ensures that $u^\mu\,u_\mu=-1$.}
\be
u^\mu
=
\sum_{b=0}^1
\Lambda^{(0)}_{\ (b)}\,e^{(b)\mu}
=
\cosh(\beta)\,e^{(0)\mu}
+
\sinh(\beta)\,e^{(1)\mu}
\ ,
\ee
where $\beta=\beta(v,r)$ is the (local) rapidity, and require that
\be
u^\mu\,\nabla_\mu
u^\nu
=
0
\ ,
\ee
so that $u^\mu$ naturally represents the 4-velocity of an (ideal) freely-falling observer.
After some straightforward calculation, we find
\be
u^\mu\,\nabla_\mu
u^\nu
=
\frac{e^{-2\,\beta}}{4}
\left[\left(2\,e^{2\,\beta}-f\right)\beta'-2\,\dot\beta+f'\right]
\left(
1,
e^{2\,\beta}+\frac{f}{2},
0,0
\right)
=
0
\ .
\ee
\par
The above equation for $\beta$ in the static regime (with $v\ge v_{\rm s}$), becomes
\be
\left(2\,e^{2\,\beta}-f\right)\beta'+f'
=
0
\ ,
\ee
where now $f=f(r)$ and $\beta=\beta(r)$.
A solution is given by
\be
\beta
=
\ln\!\left(\frac{1+\sqrt{1-f}}{\sqrt 2}\right)
\ ,
\ee
where we set an integration constant so that the rapidity is real and non-negative for $f\le 1$
(see right panel in Fig.~\ref{figMF}).
It is interesting to observe that this $\beta$ is everywhere finite for $n\ge 0$,
whereas for static states reached after the Minkowski breaking, that is for $-1<n<0$, we find 
\be
\beta
\sim
\ln\!\left(\frac{1+\sqrt{1+r^{n}}}{\sqrt 2}\right)
\sim
n\,\ln r
\to
\infty
\ ,
\ee
for $r\to 0$.
This result provides a new interpretation for the Minkowski breaking occurring at $n=0$.
\par
Using the above result in the static regime, we find
\be
u^\mu
=
\left(
-\frac{1}{1+\sqrt{1-f}},
\sqrt{1-f},0,0
\right)
\ ,
\label{u^mu}
\ee
and 
\be
T_{\mu\nu}\,u^\mu\,u^\nu
=
\rho
\ ,
\ee
where the energy density $\rho$ is still the one given in Eq.~\eqref{rhoM}
[of course, the flux $\Phi(v,r)=0$ for $v\ge v_{\rm s}$].
\subsection{Classical static end state}
We can support the conclusion that the collapse would classically end with a Schwarzschild 
singularity by studying the classical Raychaudhuri equation for the expansion~\cite{Raychaudhuri:1953yv} 
\be
\theta
=
h^{\mu\nu}\,B_{\mu\nu}
\ee
of the congruence of time-like geodesics of 4-velocity $u^\mu$ in Eq.~\eqref{u^mu}.
We recall that $B_{\mu\nu}=\nabla_\mu u_\nu$
is the gradient velocity tensor and $h_{\mu\nu}=g_{\mu\nu}+u_\mu\,u_\nu$.
The shear is also given by
\be
\sigma_{\mu\nu}
=
\frac{1}{2}\left(B_{\mu\nu}+B_{\nu\mu}\right)
-
\frac{1}{3}\,\theta\,h_{\mu\nu}
\ee
and the rotation by
\be
\omega_{\mu\nu}
=
\frac{1}{2}\left(B_{\mu\nu}-B_{\nu\mu}\right)
\ .
\ee
The Raychaudhuri equation then reads
\be
\frac{\d\theta}{\d\tau}
=
-\frac{\theta^2}{3}
-
\sigma_{\mu\nu}\,\sigma^{\mu\nu}
+
\omega_{\mu\nu}\,\omega^{\mu\nu}
-
R_{\mu\nu}\,u^\mu\,u^\nu
\ ,
\label{RayS}
\ee
where $\tau$ is the proper time along the geodesics of 4-velocity $u^\mu$.
\par
Clearly, a final configuration with $\theta=\sigma_{\mu\nu}=\omega_{\mu\nu}=0$ would be static
($\d\theta/\d\tau=0$) in the vacuum where $R_{\mu\nu}=0$.
That implies that the singular Schwarzschild geometry is a possible classical end state of the
gravitational collapse obtained in the limit $n\to-1^+$.
\par
\begin{figure}[t]
\centering
\includegraphics[height=0.4\textwidth]{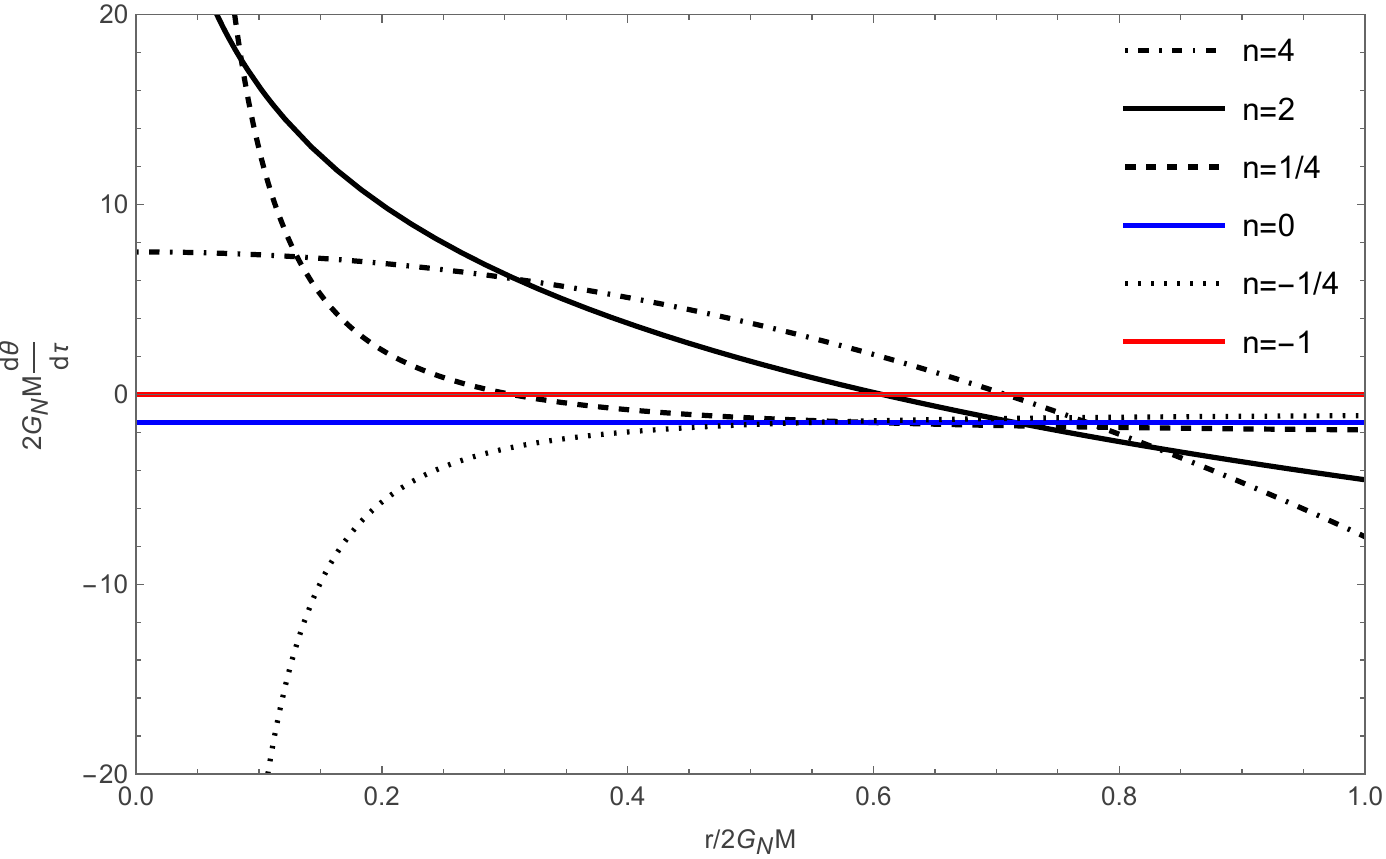}
\caption{Rate of change of the expansion~\eqref{cdtheta} for different values of $n$.}
\label{figcdtheta}
\end{figure}

For $n>-1$, we further notice that
\be
R_{\mu\nu}\,u^\mu\,u^\nu
&\!\!=\!\!&
8\,\pi\,\gn
\left(
T_{(0)(0)}
-\frac{1}{2}\,\eta_{(0)(0)}\,T^{(a)}_{\ (a)}
\right)
\nonumber
\\
&\!\!=\!\!&
4\,\pi\,\gn
\left(
\rho+\Phi+p_{\rm r}+2\,p_{\rm t}
\right)
\nonumber
\\
&\!\!=\!\!&
4\,\pi\,\gn
\left(
\Phi+2\,p_{\rm t}
\right)
\ ,
\ee
where we used the classical Einstein equations.
For any possible configuration with $\theta=\sigma_{\mu\nu}=\omega_{\mu\nu}=\Phi=0$, we then have
\be
R_{\mu\nu}\,u^\mu\,u^\nu
=
8\,\pi\,\gn\,p_{\rm t}
=
\frac{3}{\rh^2}
\left(\frac{n+1}{2-n}\right)
\left[
1-
\frac{n}{2}
\left(\frac{r}{\rh}\right)^{n-2}
\right]
\ee
and Eq.~\eqref{RayS} reads
\be
\frac{\d\theta}{\d\tau}
=
-
8\,\pi\,\gn\,p_{\rm t}
\ .
\label{cdtheta}
\ee
It follows that the congruence would start evolving from such a configuration by expanding where $p_{\rm t}<0$
and contracting where $p_{\rm t}>0$, which indeed makes the configuration unstable for all values of $n>-1$.
Some examples are shown in Fig.~\ref{figcdtheta}, from which one can see that the rate of change of the
expansion~\eqref{cdtheta} is negative everywhere for $-1<n\le 0$, that is for all configurations after the
Minkowski breaking.
This result further supports the conclusion that the classical dynamics ends in the singular
Schwarzschild geometry.
\subsection{Quantum interior}
\label{SS:quantum}
The analysis performed in the previous Sections assumes that the collapse can be effectively 
described by (semi)classical quantities, such as the mass function and the effective
energy-momentum tensor.
Quantum physics however is not always compatible with such a description.
More specifically, we observe that the classical behaviour in Eq.~\eqref{Ccond} is satisfied
by the mass function~\eqref{m-n} for $n>2$, whereas Eq.~\eqref{Qcond} and the condition
that the wavefunction $\psi=\psi(r)$ in the hydrodynamic approach be normalisable are
compatible with $n>-1$, which includes the regime of integrable singularities.
\par
\begin{figure}[t]
\centering
\includegraphics[height=0.4\textwidth]{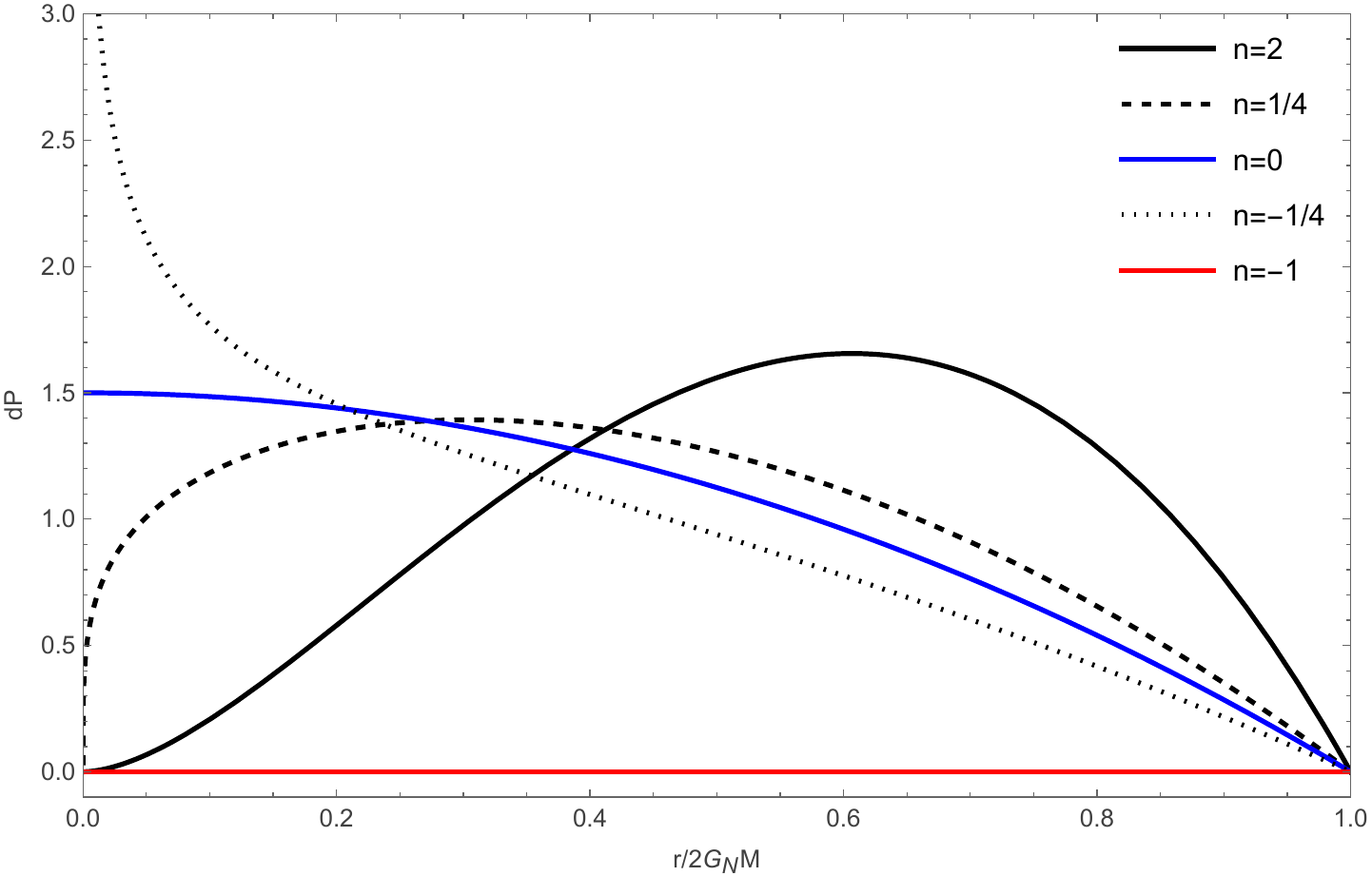}
\caption{Probability density $\d P$ in Eq.~\eqref{dP} for values of 
$-1<n< 2$ (the case $n=2$ is shown as an upper bound).
}
\label{figDP}
\end{figure}

Realistic quantum states should account for the complexity of collapsing objects
and be given by many-body wavefunctions.
As is typical for such systems, we will employ the Madelung approximation~\cite{Madelung:1927ksh}
and, in the regime $-1<n<2$ where the classical condition~\eqref{Ccond} fails,
we reverse-engineer the radial mass profile for a wavefunction at fixed $v$.
From Eq.~\eqref{Qcond} with $m=m(v,r)$ in Eq.~\eqref{m-n}, we have
\be
\mu\,|\psi|^2
=
\frac{3}{8\,\pi\,\gn\,\rh^2}
\left(\frac{n+1}{n-2}\right)
\left[
1-
\left(\frac{r}{\rh}\right)^{n-2}
\right]
\ ,
\ee
where $\mu$ has dimensions of mass and must be such that 
\be
4\,\pi
\int_0^{\rh}
r^2\,|\psi(r)|^2\,
\d r
=
1
\ .
\ee
Since 
\be
4\,\pi
\int_0^{\rh}
r^2\,\rho\,
\d r
=
M
=
\frac{\rh}{2\,\gn}
\ ,
\ee
we immediately find $\mu=M$ and
\be
|\psi|^2
=
\frac{3}{4\,\pi\,\rh^3}
\left(\frac{n+1}{n-2}\right)
\left[
1-
\left(\frac{r}{\rh}\right)^{n-2}
\right]
\ .
\label{psiR2}
\ee
In Fig.~\ref{figDP} we plot the probability density 
\be
\d P
=
4\,\pi\,r^2\,|\psi(r)|^2\,\d r
\propto
4\,\pi\,r^2\,\rho(r)\,\d r
\ ,
\label{dP}
\ee
and notice that $n=0$ marks the transition from $\d P(0)=0$ to $\d P(0)>0$
(the same behaviour shown for the density in Section~\ref{SS:sing}).
\par
\begin{figure}[t]
\centering
\includegraphics[height=0.32\textwidth]{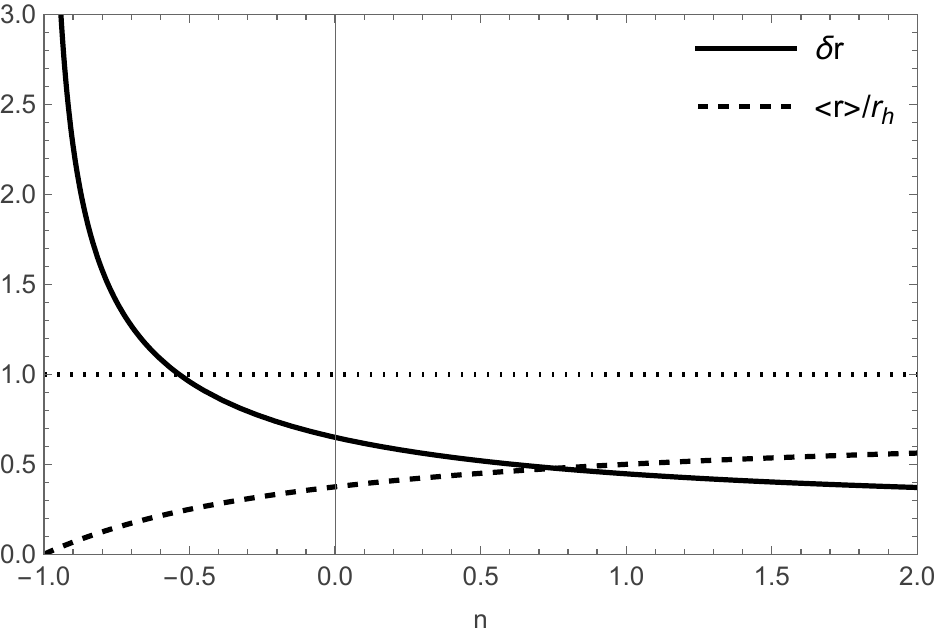}
$\quad$
\includegraphics[height=0.32\textwidth]{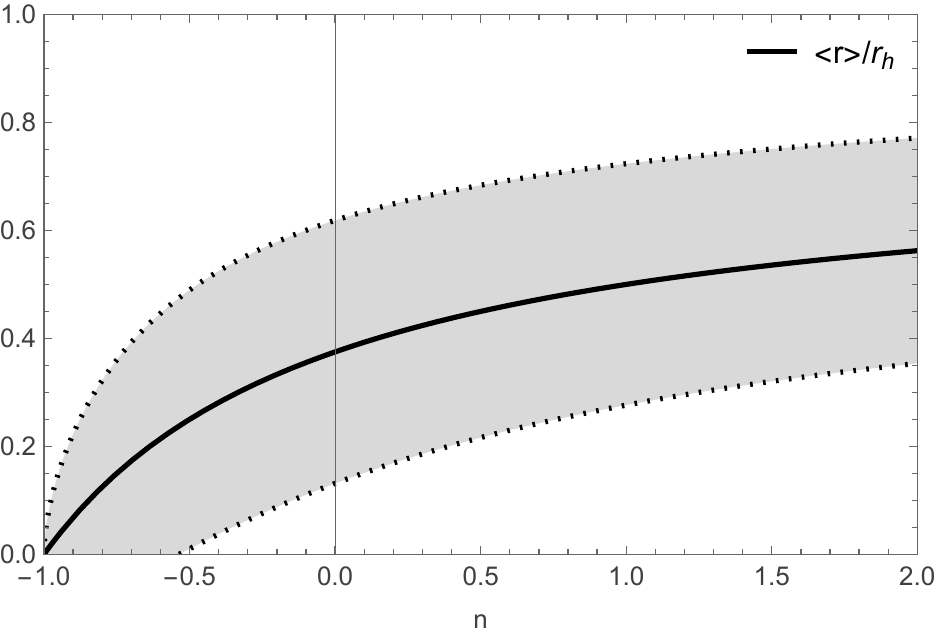}
\caption{Left panel:
expectation value $\expec{\hat r}$ in Eq.~\eqref{expr} and relative uncertainty
$\delta r$ in Eq.~\eqref{eq:dr} (dotted horizontal line corresponds to $\delta r=1$).
Right panel:
expectation value $\expec{\hat r}$ in Eq.~\eqref{expr} with shaded area covering
the strip bounded by $\expec{\hat r}\,(1\pm\delta r)$.
}
\label{figDR}
\end{figure}

We can then compute
\be
\expec{\hat r}
\equiv
4\,\pi
\int_0^{\rh}
r^3\,|\psi(r)|^2\,
\d r
=
\frac{3\,(n+1)}{4\,(n+2)}\,\rh
\label{expr}
\ee
and
\be
\expec{\hat r^2}
\equiv
4\,\pi
\int_0^{\rh}
r^4\,|\psi(r)|^2\,
\d r
=
\frac{3\,(n+1)}{5\,(n+3)}\,\rh^2
\ ,
\ee
from which we obtain the relative radial uncertainty (see left panel in Fig.~\ref{figDR})
\be
\delta r
\equiv
\frac{\Delta r}{\expec{\hat r}}
=
\sqrt{
\frac{\expec{\hat r^2}
-
\expec{\hat r}^2}
{\expec{\hat r}^2}
}
=
\sqrt{\frac{19+4\,n+n^2}{15\,(n+1)\,(n+3)}}
\ .
\label{eq:dr}
\ee
Note that the above relative uncertainty $\delta r(n=0)=\sqrt{19/45}\simeq 0.65$
and 
\be
\lim_{n\to -1^+}
\delta r
=
+\infty
\ .
\ee
The value $\delta r=1$ is reached for $n=(\sqrt{105}-14)/7\simeq -0.54$.
Moreover, 
\be
\expec{\hat r}\,(1+\delta r)
\sim
\sqrt{1+n}\,\rh
\ ,
\quad
{\rm for}\ n\to -1^+
\ ,
\ee
so that the size of the central core indeed still shrinks to zero for the Schwarzschild limit
(see right panel in Fig.~\ref{figDR}).
\subsection{Quantum static end state}
From the Madelung approximation~\cite{Madelung:1927ksh}, one expects the emergence of a quantum potential
\be
V_{\rm Q}
=
-
\frac{\hbar^2}{2\,\mu^2}\left(
\frac{\nabla^2\sqrt{\rho}}{\sqrt\rho}
\right)
=
-2\,\frac{\lp^4}{\rh^2}
\left(
\frac{\nabla^2\sqrt{|\psi|^2}}{\sqrt{|\psi|^2}}
\right)
\ ,
\ee
which yields the modified Raychaudhuri equation~\cite{Chowdhury:2020mez,Das:2013oda,Cheriyodathillathu:2026rpz}
\be
\frac{\d\theta}{\d\tau}
=
-\frac{\theta^2}{3}
-
\sigma_{\mu\nu}\,\sigma^{\mu\nu}
+
\omega_{\mu\nu}\,\omega^{\mu\nu}
-
R_{\mu\nu}\,u^\mu\,u^\nu
-
\nabla^2 V_{\rm Q}
\ .
\label{RayQ}
\ee
As in the classical case~\eqref{RayS}, we are interested in studying the stability of static final
configurations such that $\theta=\sigma_{\mu\nu}=\omega_{\mu\nu}=\Phi=0$, which now yields
\be
\frac{\d\theta}{\d\tau}
=
-
8\,\pi\,\gn\,p_{\rm t}
-
\nabla^2 V_{\rm Q}
=
0
\ .
\label{cdthetaQ}
\ee
\par
The quantum contribution,
\be
-\nabla^2 V_{\rm Q}
=
2\,\frac{\lp^4}{\rh^2}\,\nabla^2
\left(
\frac{\nabla^2\sqrt{|\psi|^2}}{\sqrt{|\psi|^2}}
\right)
\ ,
\label{DVq}
\ee
always diverges negatively for $r\to \rh$,
\be
-\nabla^2 V_{\rm Q}
\sim
-\frac{\lp^4/\rh^2}{(r-\rh)^4}
\ ,
\label{qdtheta}
\ee
which is consistent with the necessary existence of the Hawking radiation~\cite{Hawking:1975vcx}
that would reduce the ADM mass and therefore pull the radius $\rh$ towards smaller values.~\footnote{There
are studies suggesting the horizon therefore never forms (see, {\em e.g.}~Ref.~\cite{Brustein:2026kmd}).}
Note that this effect should become significant only for $r\sim\rh-\lp$.
\begin{figure}[t]
\centering
\includegraphics[height=0.4\textwidth]{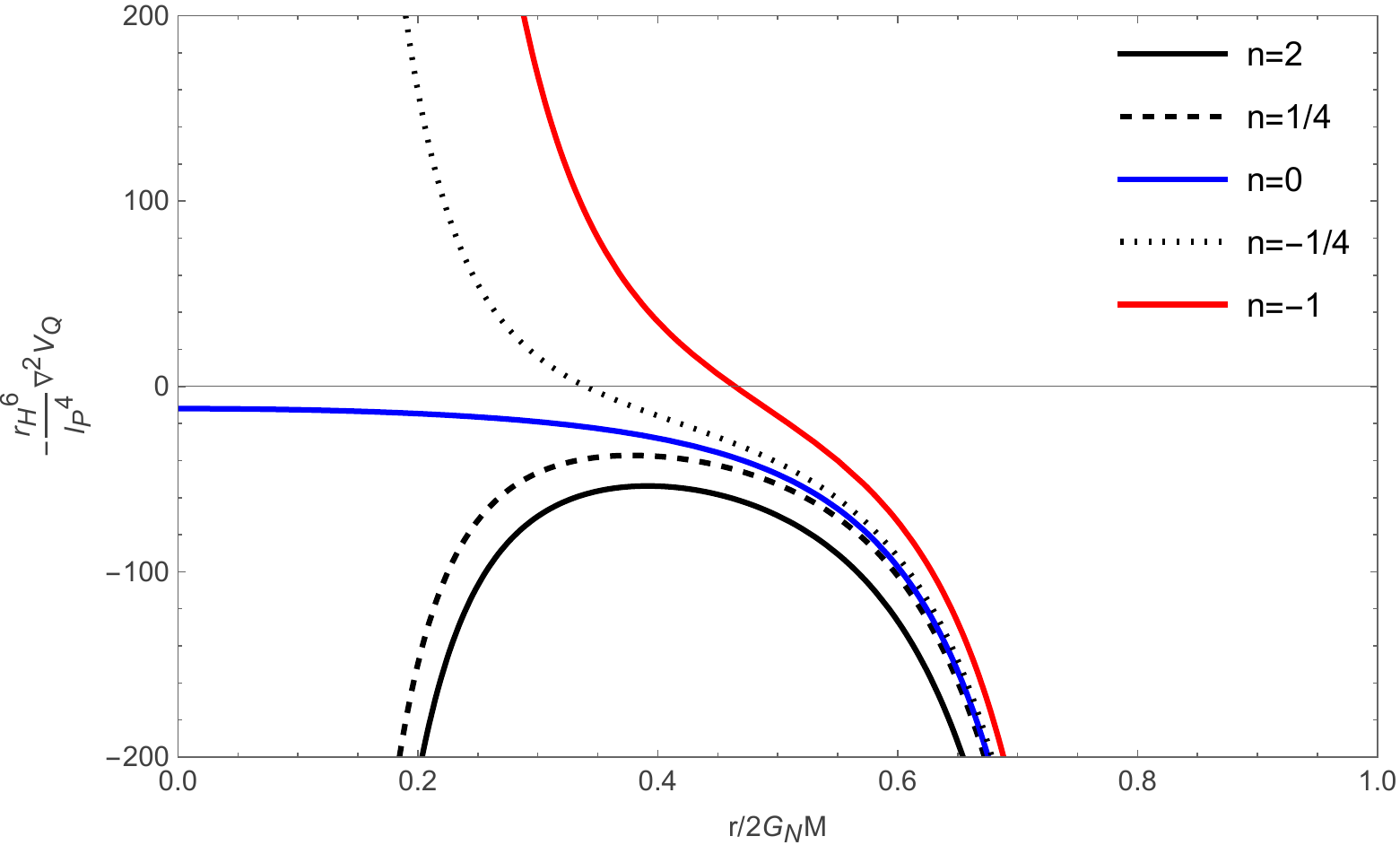}
\caption{Quantum contribution~\eqref{qdtheta} to the expansion rate for different values of $n$.
}
\label{figqdtheta}
\end{figure}
\par
Examples of the quantum contribution~\eqref{DVq} are plotted in Fig.~\ref{figqdtheta}, from which we can see that
it diverges negatively for $r\to 0$ in cases with $0<n<2$, whereas it diverges positively for $r\to 0$ in cases with $-1\le n<0$.
In fact, in the range $-1\le n<2$ and $n\neq 0$, we have
\be
-\nabla^2 V_{\rm Q}
\sim
-\frac{n\,\lp^4}{\rh^2\,r^4}
\ ,
\label{dVq0}
\ee
and $-\nabla^2 V_{\rm Q}\sim -1$ for $r\to 0$ if $n=0$.
This implies that quantum corrections would make the configurations with $0<n<2$ collapse faster near the centre,
whereas they induce a strong repulsive effect around the centre for $-1\le n<0$.
This change of behaviour precisely happens at the Minkowski breaking where $n=0$, when the Cauchy horizon
disappears and the classical contribution proportional to $p_{\rm t}$ in Eq.~\eqref{cdthetaQ} becomes everywhere negative.
However, that classical contribution vanishes for $n\to-1^+$, which means that the positive quantum contribution
around the centre given in Eq.~\eqref{dVq0} should dominate after the Minkowski breaking and stop the evolution before the
Schwarzschild singularity forms.
\par
Note further that the change of sign in $\nabla^2 V_{\rm Q}$ occurs around $r\simeq \rh/2$ for all values of $-1\le n<0$,
which points to the possible formation of a quantum core of significant size at the end of the gravitational collapse.
However, definite conclusions will require solving for the time evolution from the complete Raychaudhuri Eq.~\eqref{RayS}
with the addition of the quantum contribution~\eqref{DVq}, which is left for future (numerical) studies.
\section{Conclusions and outlook}
\setcounter{equation}{0}
\label{S:conc}
We have analysed the models of gravitational collapse involving integrable singularities put forward in
Refs.~\cite{Ovalle:2025pue,Ovalle:2026lxb,Ovalle:2024wtv} from a semiclassical and quantum perspectives.
\par
In the semiclassical approximation, we have confirmed that the 
effective energy-momentum tensor for the interior of the collapsing
object contains a non-vanishing flux term during the time evolution,
and can undergo a transition from being regular to containing a singularity
of the integrable type in the centre.
Moreover, the flux of energy remains finite in the centre only if the Schwarzschild limit 
is approached asymptotically.
\par
We next analysed the integrable stage of the collapse in the Madelung
approximation for the wavefunction of collapsing matter in possible end states
reached before the Schwarzschild singularity forms and showed that the quantum correction
to the Raychaudhuri equation starts to work against the formation of the Schwarzschild 
singularity exactly after the ``Minkowski breaking'', when the Cauchy horizon disappears.
\par
We can conclude that several effects seem to occur at the ``Minkowski breaking'',
when the function $n=n(v)$ crosses zero from positive to negative values:
the Cauchy horizon disappears;
the derivative of the MSH mass function at the origin suddenly goes from zero to infinity; 
the metric function $f$ correspondingly goes from one to minus infinity;
the quantum contribution to the Raychaudhuri equation for the expansion
of the collapsing fluid jumps from minus infinity to plus infinity at the centre.
Overall, the Minkowski breaking appears as a rather discontinuous transition
in the evolution of integrable singularities and, as such, it deserves further 
investigation. 
\par
It is important to remark that here we have focused on finding results that should not
strongly depend on the actual time evolution of the MSH mass function~\eqref{m-n}, hence 
the function $n=n(v)$.
In order to understand the actual role of the Minkowski breaking (and Hawking evaporation),
a more detailed modelling of the collapse is necessary, which will have to be performed by solving
dynamical equations, most likely numerically, in future works.
\section*{Acknowledgments}
R.C.,~A.G.~and A.K.~are partially supported by the INFN grant FLAG. 
A.G.~is supported by the Italian Ministry of Universities and Research (MUR)
through the ``BACHQ: black holes and the quantum'' (Grant no.~J33C24003220006).
J.O.~was partially supported by ANID-FONDECYT Grant No.~1250227.
The work of R.C.~and A.G.~has also been carried out in the framework of activities
of the National Group of Mathematical Physics (GNFM, INdAM).
\end{document}